\documentclass[sigconf,nonacm]{aamas}

\usepackage{balance}

\setcopyright{none}
\AtBeginDocument{%
  \fancypagestyle{firstpagestyle}{\fancyhf{}\fancyhead[C]{\sffamily\itshape\fontsize{16}{18}\selectfont Under peer-review at AAMAS 2027}\fancyfoot[C]{\footnotesize\thepage}}%
  \fancypagestyle{standardpagestyle}{\fancyhf{}\fancyfoot[C]{\footnotesize\thepage}}%
  \pagestyle{standardpagestyle}}

\title[Centralized Learning, Embodied Coordination, and Tabular Studies]{Does Centralized Learning Help Embodied Agents Coordinate, and When Can a Tabular Study Tell?}

\author{Muhammad Ahmed Atif}
\affiliation{
  \institution{Dhanani School of Science and Engineering, Habib University}
  \city{Karachi}
  \country{Pakistan}}
\email{m.ahmedatif720@gmail.com}

\author{Nehal Naeem Hajji}
\affiliation{
  \institution{Dhanani School of Science and Engineering, Habib University}
  \city{Karachi}
  \country{Pakistan}}
\email{nehalnaeem13@gmail.com}

\author{Muhammad Ebad Atif}
\affiliation{
  \institution{Dhanani School of Science and Engineering, Habib University}
  \city{Karachi}
  \country{Pakistan}}
\email{ma09639@st.habib.edu.pk}

\author{Mohammad Shahid Shaikh}
\affiliation{
  \institution{Dhanani School of Science and Engineering, Habib University}
  \city{Karachi}
  \country{Pakistan}}
\email{shahid.shaikh@sse.habib.edu.pk}

\begin{abstract}
Tabular studies are a common way to compare independent and centralized multi-agent learners, because they avoid function approximation. A natural expectation is that centralized learning, which learns joint actions, helps agents coordinate, especially when their bodies make solo success hard. Earlier work tests this only in parts: it measures coordination for one learning method, compares methods with the agents' bodies fixed, or varies speed but reports only how well the team did. We test it in a predator--prey environment, where teammates cooperatively hunt together while two teams compete. We compare independent Q-learning (IQL) with central Q-learning (CQL) in all four predator and prey pairings, under limits on speed and stamina, on a 5$\times$5 and an 8$\times$8 grid trained for 200,000 episodes each, and check that the agents learn on both. We measure coordination directly, with predictions for the main measurement fixed in advance. In this task, where one predator can capture prey alone, we find three things. First, CQL costs its agents in both roles on both grids, and its predators coordinate no more than IQL's. Second, its one benefit comes from acting on the best joint action: this cut capture time by 42\% on the small grid with faster prey and unlimited stamina, but helped nowhere on the larger grid, where the best joint action had mostly never been tried. Third, on the larger grid a tie-breaking default decided which method won, even though the agents had passed a learning check. A tabular comparison can therefore be trusted only if it also checks how much of the state space the agents covered and implements the methods alike. We propose simple checks for both.
\end{abstract}


\begin{document}

\maketitle

\section{Introduction}
\label{sec:intro}

A team of learning agents can be trained in two basic ways. Independent learners each learn on their own and treat their teammates as part of the environment. A centralized learner learns values for joint actions, the combinations of its members' actions \cite{matignon2012independent,albrecht2024marl_book}. Which approach works better, and why, is a long-standing question. With deep neural networks, independent learners are often competitive \cite{papoudakis2021benchmarking}. Tabular learners, which store one value for every state and action, remain a common way to compare the two. Without function approximation, a difference in performance can be traced to the learning method itself \cite{tan1993multi,claus1998dynamics}. Predator--prey games are a standard test bed for such comparisons: teammates hunt the same prey while the two teams compete \cite{benda1986optimal,tan1993multi,stone2000multiagent,lowe2017multiagent}.

When agents have bodies, those bodies limit what each agent can do alone. How fast hunters move relative to their prey affects whether they succeed together \cite{desouza2021decentralized,tsutsui2024collaborative}. Physical differences between agents change how teams are best trained \cite{bettini2023heterogeneous}. In animals, too, the energy cost of a chase shapes how predators hunt \cite{hubel2016energy,wilson2018biomechanics}.

A natural expectation follows. Centralized learning represents what the team does jointly, so it should help agents coordinate, most of all when their bodies make it hard for one agent to succeed alone. This reasoning motivated joint-action learning \cite{claus1998dynamics,guestrin2002coordinated}, although early tabular results were mixed \cite{claus1998dynamics,kok2004sparse}. We test the expectation directly: does centralized learning help embodied agents coordinate?

Earlier work tests parts of it. Some studies measure coordination for a single learning method \cite{barton2019measuring,desouza2021decentralized,heuillet2022collective,tsutsui2024collaborative}. Others compare learning methods with the agents' bodies held fixed \cite{yong2009coevolution,papoudakis2021benchmarking,cardei2026coordination}, or vary speed but report only how well the team did, such as capture time \cite{lowe2017multiagent,huttenrauch2019deep,xu2025emergent}. Coordination is often judged from performance, although a higher reward alone does not show that agents coordinate \cite{barton2019measuring,biswas2026helping}. And to our knowledge, no study separates what a centralized learner learns from how it acts on what it learned.

Two further issues bear on any comparison of learning methods. Testing trained agents against random play is recommended practice \cite{tan1993multi,bellemare2013arcade,machado2018revisiting,patterson2024empirical}, yet the evaluation protocol that Gorsane et al. propose for cooperative multi-agent learning includes no such check \cite{gorsane2022towards}. And small implementation details can decide which method comes out ahead \cite{henderson2018deep,tucker2018mirage,engstrom2020implementation,sun2020reevaluation}. One such detail is how a greedy learner breaks ties between actions it values equally, which Section 6 shows can decide a comparison. This raises a second question: when can a tabular comparison answer the first, our main question? We ask it because the answer to the first is only as good as the comparison that gives it.

We bring these parts together in one controlled study. Two predators chase two prey on a grid under three speed settings (equal speed, faster predators, faster prey) and two stamina settings (unlimited and limited). One predator can capture a prey alone, so the task allows teamwork but does not require it. We compare independent Q-learning (IQL) with central Q-learning (CQL), which learns one table over the team's joint actions. We pair the two methods in all four ways, so that each method plays each role. CQL differs from IQL in two ways, so we add two variants to tell them apart. First, CQL learns from the team's summed reward, while each IQL agent learns from its own reward. The first variant, IQL with team reward, shows the effect of sharing the reward on its own. Second, each CQL agent acts by averaging the joint table over its partner's possible actions. In the second variant, joint-execution CQL, both agents instead take the best joint action, which shows whether acting on the joint action makes a difference.

We train on two grids for the same 200,000 episodes. They differ in how much experience the agents get. On a 5$\times$5 grid the agents visit nearly every possible state (every arrangement of the four agents) many times. On an 8$\times$8 grid they reach only about half of them, most only once or twice.

Our contributions are as follows.

\begin{enumerate}
\item A controlled design that crosses the learning method, the agents' bodies and all four pairings of predator and prey methods, and measures coordination directly against random baselines, with predictions fixed before the main measurement (Sections 3 and 4).
\item A test of when centralized learning helps coordination. It separates what CQL learns, a joint table and a shared reward, from how it acts: each agent averaging over its partner's actions, or both acting on the best joint action (Section 5).
\item The same comparison on two grid sizes at one training budget, showing how its conclusions depend on how much of the state and joint-action space the agents explored (Sections 5 and 6).
\item Simple checks for whether a tabular comparison can be trusted: a learning check against random opponents, a measure of state-space coverage and matched tie-breaking between equally valued actions. They build on established advice (Section 6).
\end{enumerate}

\section{Related Work}
\label{sec:related}

Tan compared independent and cooperating tabular learners in this same hunter--prey task and noted that the number of joint states and actions grows exponentially with the number of agents \cite{tan1993multi}. Claus and Boutilier studied joint-action learners, in which each agent keeps its own table over joint actions. These often performed little differently from independent learners, because a joint learner's value for its own action, averaged over its belief about the partner, is close to what an independent learner estimates \cite{claus1998dynamics}. With deep networks, independent learners are often competitive \cite{dewitt2020independent,papoudakis2021benchmarking,yu2022surprising,lyu2021contrasting}. Methods also differ in how they act on joint values. In its textbook form, central Q-learning keeps one table for the whole team and acts on the best joint action \cite{albrecht2024marl_book}. Joint-action learners weight joint values by a learned belief about the partner. Claus and Boutilier compared several such rules, but only in repeated matrix games \cite{claus1998dynamics}. Value-factorization methods, which build the team's value from one value per agent, ask when each agent's greedy action matches the greedy joint action \cite{son2019qtran,wang2021towards}. To our knowledge, no study has taken one centralized learner and compared acting on the best joint action with averaging over the partner in a task that unfolds over many steps. If the way a centralized learner acts decides whether it helps, a comparison that fixes one way of acting may judge the method unhelpful when only that way of acting is.

Predator--prey benchmarks build in speed differences as fixed design choices \cite{lowe2017multiagent,huttenrauch2019deep,terry2021pettingzoo}. Studies that vary relative speed use a single learning method each and find that speed shapes whether hunters succeed together \cite{desouza2021decentralized,tsutsui2024collaborative,xu2025emergent}. Work on teams of different robots treats speed as a physical difference but leaves out limits that change during a task, such as battery level \cite{bettini2023heterogeneous}. Elsewhere, energy costs appear only in simulators \cite{noda1998soccer,kurach2020google} and animal studies \cite{hubel2016energy,wilson2018biomechanics}. The one study we know of that crosses the learning method with the agents' bodies is by Lowe et al. They paired deep learners of different methods across the two roles at two prey speeds and reported only how well the teams did \cite{lowe2017multiagent}. How a stamina budget affects coordination between learning agents appears not to have been studied.

Coordination can be measured directly. Studies count how often several predators take part in a capture \cite{leibo2017sequential}, how agents divide targets \cite{gorelick2004normalized,liu2022motor} and whether they encircle prey \cite{desouza2021decentralized}. Others measure how the agents' actions depend on each other \cite{jaques2019social}. Still others measure how much a team loses when a member is replaced by one from another run \cite{lanctot2017unified,hu2020otherplay,barrett2011empirical} or by a random agent \cite{heuillet2022collective}. Each measure can mislead on its own. A higher reward does not necessarily mean more coordination \cite{barton2019measuring,biswas2026helping}. Encirclement can arise from simple rules followed independently \cite{muro2011wolfpack}. And agents that coordinate through the environment can show little dependence between their actions \cite{yong2009coevolution}. That dependence must in any case be measured within the same situation \cite{lowe2019pitfalls}. Such measures have been applied to one learning method \cite{barton2019measuring,tsutsui2024collaborative} or across methods with the bodies fixed \cite{yong2009coevolution,cardei2026coordination}. As far as we can tell, they have never been applied across learning methods and bodies together, and the deep-learning comparisons above report only how well the teams did. Yet from performance alone, a method that fails to coordinate cannot be told apart from one that is simply worse at the task.

Greedy action selection picks the highest-valued action, so it must break ties between equally valued actions. Textbooks leave the rule open, to be made ``in some arbitrary way, perhaps randomly'' \cite{suttonbarto2018rl}. When values start at zero and rewards are negative, untried actions look best \cite{suttonbarto2018rl} and tie with each other. Potential-based reward shaping, a common way to guide learning with extra rewards, is equivalent to a starting value that is the same for every action in a state \cite{wiewiora2003potential}, so it cannot break those ties. A tie rule shared by all agents works as a convention that coordinates them \cite{boutilier1996planning,busoniu2008comprehensive}, a device that ``will only work in self-play'' \cite{wei2016lenient}. Different tie rules also lead a tabular learner to different policies \cite{vamplew2024issues}. The literature treats the tie rule as a choice within one algorithm. It does not consider a rule that differs between the methods being compared. More broadly, implementation details left out of papers can account for most of a method's reported advantage \cite{engstrom2020implementation,tucker2018mirage,hu2023riit}. Checking trained agents against random play and training until performance levels off are recommended practice, and Tan did both in this task \cite{tan1993multi,bellemare2013arcade,machado2018revisiting,patterson2024empirical}. A leading evaluation protocol for cooperative multi-agent learning nonetheless compares methods only with each other, with no random baseline and no check on state-space coverage \cite{gorsane2022towards}. So a difference that passes significance tests can come from an implementation default or from thin experience, and such a protocol would not catch it.

Taken together, the pieces are in place, but no study we know of combines them. We measure directly how relative speed and a stamina budget change coordination under independent and central Q-learning, across all four pairings. We also check that the agents learned, how much of the state space they covered and that the two methods are implemented alike. Whether centralized learning helps embodied agents coordinate can only be answered by measuring coordination between agents that have learned, under the limits meant to make it matter.

\section{Setting}
\label{sec:setting}

\subsection{Environment}

The environment is an open-source predator--prey gridworld for multi-agent reinforcement learning \cite{anon2025gridworld}. Two predators and two prey move on a 5$\times$5 or 8$\times$8 grid with a few obstacles (2 and 6 cells), placed differently for each seed. At every step, all agents choose at once among five actions: a move in one of four directions or staying put. An agent with speed 2 moves up to two cells, unless a wall or obstacle stops it. Agents may share a cell. A prey that shares a cell with a predator after all agents have moved is captured but stays in play. An episode ends after two captures in total, or after 200 steps.

We vary the agents' bodies in two ways. Speed has three settings: equal speed, faster predators (speed 2 against 1) and faster prey (1 against 2). Stamina is unlimited or limited. With limited stamina, a move costs stamina equal to the agent's speed. Staying put, or trying a move the agent cannot afford, regains one unit, up to five. Together these give six settings.

A state is the positions of the four agents. Since agents never stand on obstacles, this gives about 280,000 possible states on 5$\times$5 and 11.3 million on 8$\times$8. Stamina is not part of the state, so the limited-stamina settings are partially observable: an agent cannot see how much stamina it has left. A predator receives +100 in each step in which it shares a capture cell, every predator receives $-$5 per step, and a captured prey receives $-$100. To guide learning we also use potential-based reward shaping \cite{wiewiora2003potential}: the learning update adds $F = \gamma\Phi(s') - \Phi(s)$, where $\gamma$ is the discount and $\Phi$ is minus a predator's distance to the nearest prey, or a prey's distance to the nearest predator. Methods that learn from the team's reward sum their members' $F$.

\subsection{Learning methods}

All four methods are versions of Q-learning, and they differ only in what they learn and how they act. A Q-learner keeps a table with one value for each state and action and mostly takes the greedy action, the one with the highest value. In independent Q-learning (IQL), each agent learns its own table over its own five actions, from its own reward. Central Q-learning (CQL) keeps one table per team over the team's 25 joint actions and learns from the team's summed reward and shaping. IQL with team reward is IQL trained on CQL's signal. It separates the effect of a shared reward, a matter of credit assignment, from that of a joint table.

Each CQL agent acts by averaging the joint table over its partner's actions and taking the best result, and explores on its own. This is the value of an individual action that Claus and Boutilier's joint-action learners use, with a fixed, uniform belief about the partner \cite{claus1998dynamics}. The textbook version of central Q-learning instead acts on the best joint action \cite{albrecht2024marl_book}. We use the averaging rule because our implementation used it from the start. To test the alternative, we add joint-execution CQL. It has the same table and learning update, but both agents act on the best joint action and the team explores as one. Predators and prey use IQL or CQL in all four pairings. We call a pairing self-play when both teams use the same method, and cross-play when they differ. IQL with team reward ran only in self-play, and joint-execution CQL only as predators against CQL prey.

All methods start from zero values and share one set of learning parameters (step size 0.25, discount 0.9), which we did not tune for each method. Exploration is $\varepsilon$-greedy: an agent takes a random action with probability $\varepsilon$, which starts at 1 and shrinks by a factor of 0.99 every 100 episodes, down to 0.1. When several actions share the highest value, every method breaks the tie at random. Sections 5 and 6 examine what happens when one does not. In all methods, the learning update uses the next state's value at the last step of an episode. This is correct when the 200-step limit cuts an episode short, but not after the final capture, where it should be zero.

\subsection{Training and evaluation}

On both grids, agents train for 200,000 episodes. We measure performance by capture time: the mean episode length over the last 10,000 training episodes, that is, the time the predators take to make two captures (at most 200 steps). Shorter is better for the predators. To check that the agents learned, we replay their final tables against opponents that move at random, in both roles, for 1,000 episodes per seed. We measure coordination in separate replays with learning switched off (Section 4). Section 5.4 reports three controls: training without shaping, longer training and a best-response test.

Every main condition has 10 seeds. A seed fixes the obstacle layout and the random numbers, so conditions with the same seed can be compared in pairs. We report the mean paired difference with a 95\% confidence interval and test it with an exact two-sided Wilcoxon signed-rank test. With 10 seeds, its smallest possible p-value is 0.002. Within each family of tests, Holm's method corrects the p-values for multiple testing. Rerunning a condition with the same seed reproduces its results exactly.

\section{Measuring Coordination}
\label{sec:measures}

\subsection{What we call coordination}

As Section 2 noted, capturing prey faster does not by itself show coordination, and single measures also mislead \cite{barton2019measuring,biswas2026helping,muro2011wolfpack}. Agents can even coordinate through the environment while their moves look unrelated, each reacting to changes the other makes \cite{yong2009coevolution,lowe2019pitfalls}. We therefore use two names. \emph{The pair beats its parts} when the two predators capture more together than each would with a random partner. In other words, they gain from hunting together. A team \emph{coordinates} in a setting when the pair beats its parts and its predators also catch prey together or flank them more often than random predators do. We report every measure, but no single measure counts as coordination on its own.

\subsection{Replays of the trained agents}

Ongoing learning distorts measures of how agents depend on each other \cite{papadopoulos2026nonstationarity}. We therefore measure coordination on the final tables with learning switched off. Each replay runs 500 episodes in the seed's own obstacle layout. Trained agents take a random move 10\% of the time, as they did at the end of training. As a check, we repeat each replay with agents that always take their best move. We compare every predator measure with random predators that face the same trained prey. This way we do not count the prey's skill as the predators' coordination. In the same way, we compare the prey measure with random prey that face the same trained predators. The pair and partner measures also replay each trained predator with a random partner and with a partner trained on another seed. On both grids we replay all four pairings of IQL and CQL in all six settings. We also replay joint-execution CQL predators against CQL prey at equal speed and with faster prey, under both stamina settings.

\subsection{Measures}

Table~\ref{tab:measures} defines the seven measures. We count distances in grid steps (horizontal plus vertical). Move dependence did not work as planned. We meant it to tell coordination through moves from coordination through the environment, but it rises whenever both predators react to the same situation, even when neither responds to the other. We report it without that label, which changes no verdict.

\begin{table}[t]
\caption{Coordination measures. We compare predator measures with random predators facing the same trained prey, and prey spacing with random prey facing the same trained predators.}
\label{tab:measures}
\small
\begin{tabular}{@{}p{0.27\columnwidth}p{0.68\columnwidth}@{}}
\toprule
Measure & What it measures \\
\midrule
Splitting up & Share of steps in which the two predators' nearest prey differ \cite{gorelick2004normalized,liu2022motor} \\
Catching together & Share of captures with both predators within one step of the captured prey \cite{leibo2017sequential} \\
Flanking & Among steps with both predators one or two steps from the same prey: the share in which they are on opposite sides of it \cite{desouza2021decentralized} \\
Move dependence & How much one predator's move tells about the other's move in the same situation (mutual information in bits, minus a shuffled baseline) \cite{jaques2019social,lowe2019pitfalls} \\
Partner swap & Extra capture time with a partner trained on another seed or with a random partner \cite{barrett2011empirical,carroll2019utility,strouse2021collaborating} \\
Pair beats its parts & The pair's captures per step above two random predators, minus the same gain for each member hunting with a random partner. Zero for two independent hunters \cite{heuillet2022collective} \\
Prey spacing & Mean distance between the two prey and the share of steps in which they are at least 3 steps apart (4 on 8$\times$8) \cite{hahn2019emergent} \\
\bottomrule
\end{tabular}
\end{table}

\subsection{Reading the results}

In the results, a difference counts only when three conditions hold. First, it goes the same way in at least 8 of 10 seeds. Second, an exact paired Wilcoxon test gives p < 0.05. Third, it reaches a minimum size: 5 percentage points for shares, 0.05 bits for move dependence and 5\% of the team's capture rate for the pair measure. Holm's correction applies within each prediction, for example over the 24 combinations of pairing and setting for the prediction of little coordination on 5$\times$5. A script scores each prediction automatically as supported, refuted (when the opposite direction passes) or not supported. We report every verdict.

\subsection{Predictions}

Before computing any coordination measure on trained agents, we wrote down twelve predictions and these reading rules and committed them with time stamps. Our main prediction was that trained predators coordinate little in this task, whatever their method. We also predicted that they coordinate more when slower than the prey, as found in collaborative hunting \cite{tsutsui2024collaborative}. And we predicted that they coordinate more when stamina is limited. Following the natural expectation of Section 1, we predicted that CQL predators coordinate more than IQL predators. A competing prediction held that they coordinate no more. A further prediction held that joint-execution CQL predators score higher than CQL predators on the coordination measures when the prey are faster.

Table~\ref{tab:verdicts} in Section 5 gives the verdicts on the 5$\times$5 grid. This was the main measurement. The 8$\times$8 coordination measures (Section 5.3) are exploratory. They use the same measures and rules, but we wrote no prediction for them. We do not report one of the twelve predictions, which concerned the 8$\times$8 grid at an earlier training budget.

\section{Results}
\label{sec:results}

\subsection{The agents learn}

Before comparing the methods, we check that the agents learned by playing them against opponents that move at random. In all six settings on both grids, trained IQL and CQL agents beat random ones in all 10 seeds. On 5$\times$5, trained predators caught random prey 1.8 to 2.9 times faster than random predators, and trained prey survived random predators 2.4 to 6.0 times longer than random prey. On 8$\times$8 the gains were smaller, 1.14 to 1.45 times for predators and 1.26 to 3.21 times for prey, but present everywhere.

The 8$\times$8 agents learned less, as they saw less (Table~\ref{tab:coverage}). On 5$\times$5 they visited almost every possible state many times. On 8$\times$8 they visited about half, and about half of those only once. CQL also tried a much smaller share of its joint actions than IQL tried of its single actions. The differences that follow are therefore between agents that learned at this budget, though less so on the larger grid.

\begin{table}[t]
\caption{How much of the state and action space the predators tried, ranges over the six settings (means over 10 seeds). Untried greedy action: share of the visits to a state in which the action the table rated best had never been taken.}
\label{tab:coverage}
\small
\begin{tabular}{@{}lll@{}}
\toprule
 & 5$\times$5 & 8$\times$8 \\
\midrule
Possible states visited & 92--99\% & 40--58\% \\
Visits per visited state & 11--27 & 1.7--2.8 \\
Actions tried per state, IQL & 67--83\% & 34--43\% \\
Joint actions tried per state, CQL & 18--27\% & 7--9\% \\
Untried greedy action, IQL & 2--7\% & 51--82\% \\
Untried greedy action, CQL & 7--27\% & 87--94\% \\
\bottomrule
\end{tabular}
\end{table}

\subsection{Performance}

\begin{figure}[t]
\centering
\includegraphics[width=\columnwidth]{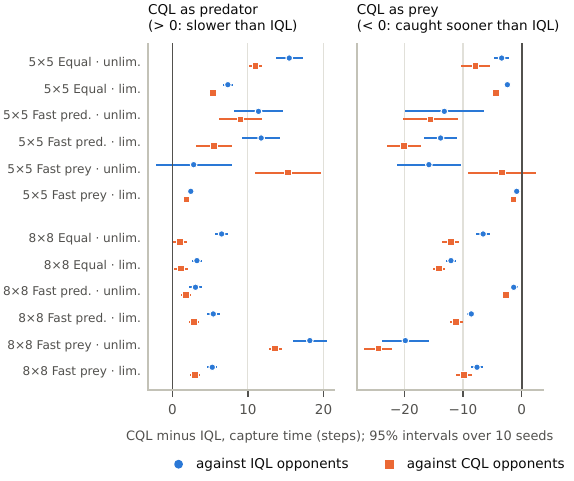}
\caption{CQL minus IQL in capture time, for each role and each opponent method, in the six settings, on both grids (means and 95\% intervals over 10 seeds, paired by seed). On the left, positive values mean CQL predators are slower; on the right, negative values mean CQL prey are caught sooner.}
\label{fig:roles}
\end{figure}

CQL performed worse than IQL in both roles on both grids (Figure~\ref{fig:roles}). Each method played each role against both types of opponent, giving 24 comparisons per grid. On 5$\times$5, 22 of the 24 were significant after Holm correction. The two exceptions, both with faster prey and unlimited stamina, were not significant. On 8$\times$8 all 24 were. CQL predators took up to 18.2 steps longer to capture, and CQL prey were caught up to 24.4 steps sooner. The costs also appear against opponents that do not learn. In the learning check of Section 5.1, CQL agents gained less over random play than IQL agents in all 12 comparisons on 8$\times$8 and in 9 of 12 on 5$\times$5, where the three exceptions all had faster prey. The four pairings show the same pattern. In every grid and setting, IQL predators against CQL prey gave the shortest capture times, and CQL predators against IQL prey the longest.

In self-play the two costs mix. A simple rule decides which method self-play favours: the larger cost wins. In all 12 combinations of grid and setting, the cost of CQL predators plus the cost of CQL prey, each measured against IQL opponents, had the same sign as the self-play difference. On 5$\times$5 the larger cost depended on speed, so speed reversed which method looked better. At equal speed the predators' cost was larger, and CQL self-play took longer than IQL self-play. With faster predators the prey's cost was larger, and CQL self-play took less time, significantly so with limited stamina. On 8$\times$8 the prey's cost was larger in five of six settings, so there was no reversal. CQL self-play was shorter in those five and level in the sixth. Table~\ref{tab:ties} (matched columns) gives all twelve differences. The sizes do not match exactly, because in self-play the two teams also adapt to each other.

Where does CQL's cost in self-play come from? IQL with team reward splits it into two parts: the shared reward, which raises a credit assignment problem, and the joint table. On 5$\times$5 at equal speed, the shared reward lengthened self-play episodes by 3.3 steps, and the joint table added another 4.3. On 8$\times$8 the joint table made no measurable difference in five of six settings. The shared reward alone shortened episodes by 4.9 to 10.9 steps in five settings. Finally, the tie rule made no difference on 5$\times$5: breaking ties by the first action rather than at random changed capture time by at most 2 steps for either method. On 8$\times$8 it still mattered (Section 6).

\subsection{Coordination}

On 5$\times$5 the measures test predictions fixed in advance (Table~\ref{tab:verdicts}). On 8$\times$8 they are exploratory (Section 4).

\begin{table}[t]
\caption{Predictions on the 5$\times$5 grid, fixed before measurement, and their verdicts. Scored with 10\% random moves, as in training, and checked with best moves only. A prediction is refuted when the opposite direction passes our rule.}
\label{tab:verdicts}
\small
\begin{tabular}{@{}p{0.56\columnwidth}p{0.2\columnwidth}p{0.17\columnwidth}@{}}
\toprule
We predicted that \ldots & Verdict & Best moves \\
\midrule
trained predators coordinate little, whatever their method & refuted (narrowly) & supported \\
predators coordinate more when slower than the prey & refuted & refuted \\
predators coordinate more with limited stamina & refuted & refuted \\
CQL predators coordinate more than IQL predators, yet capture more slowly & partly (performance only) & partly \\
CQL predators coordinate no more than IQL predators & not supported & not supported \\
joint-execution CQL predators score higher than CQL predators on the coordination measures when the prey are faster & supported & supported \\
speed and stamina change coordination differently for IQL and CQL & supported & supported \\
the prey's method changes how much the predators coordinate & supported & supported \\
IQL predators depend more on their usual partner than CQL predators & not supported & mixed \\
faster prey keep further apart & refuted & refuted \\
the reversal of the method ranking comes without a change in coordination & refuted & refuted \\
\bottomrule
\end{tabular}
\end{table}

\begin{figure*}[t]
\centering
\includegraphics[width=\textwidth]{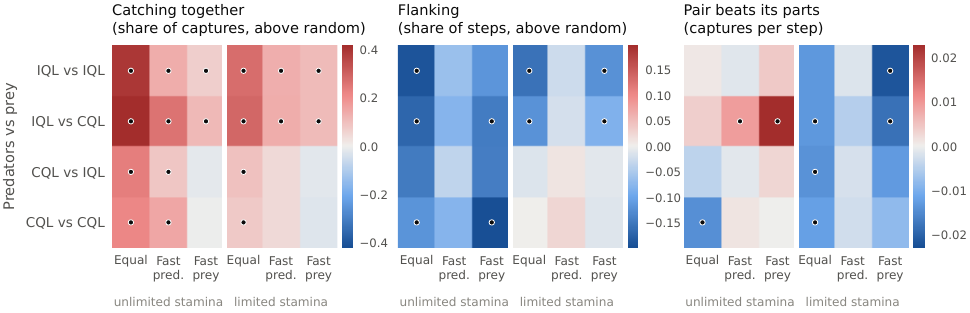}
\caption{5$\times$5 coordination measures, four pairings (predators against prey) in six settings, as differences from random predators facing the same trained prey; the pair measure is zero for independent hunters. Red: above; blue: below. Dots mark cells that pass our rule (Section~\ref{sec:measures}) in either direction, with Holm correction over the 24 cells.}
\label{fig:heatmaps}
\end{figure*}

Trained predators ganged up without coordination (Figure~\ref{fig:heatmaps}). On 5$\times$5, compared with random predators facing the same trained prey, they caught prey together more often in 20 of 24 pairing and setting combinations. Yet they flanked less in 21 of 24, mostly coming at the prey from the same side, and only twice did the pair beat its parts. The only coordination on this grid came from IQL predators against CQL prey with unlimited stamina, in two settings: faster predators and faster prey. Both cases were fragile. They only just passed the corrected threshold and disappeared when the agents took their best moves only. On 8$\times$8 there was none. In none of the 24 combinations did catching together, flanking or the pair's gain rise above random predators under our rule.

CQL predators coordinated no more than IQL predators. On 5$\times$5 they caught prey together less often in all 12 comparisons, so the prediction of no difference was not supported (Table~\ref{tab:verdicts}). In none were they more coordinated by our definition. On 8$\times$8 the two methods did not differ in any comparison, since neither coordinated. Table~\ref{tab:effects} gives effect sizes and intervals for these contrasts.

\begin{table}[t]
\caption{Coordination effect sizes (5$\times$5 unless noted): mean paired difference over 10 seeds [95\% CI], seeds in its direction, Holm-corrected Wilcoxon $p$. Shares are excess over random predators.}
\label{tab:effects}
\footnotesize
\begin{tabular}{@{}lrrr@{}}
\toprule
Contrast & Mean [95\% CI] & Seeds & $p$ \\
\midrule
Catching together, largest cell & +0.42 [0.38, 0.45] & 10/10 & .047 \\
Pair's gain, faster predators$^{a}$ & +0.009 [0.005, 0.013] & 10/10 & .047 \\
Pair's gain, faster prey$^{a}$ & +0.023 [0.014, 0.032] & 10/10 & .047 \\
Catching together, CQL $-$ IQL & $-$0.05 to $-$0.21 & $\geq$9/10 & .023 \\
Joint execution $-$ CQL, pair's gain$^{b}$ & +0.035 [0.030, 0.040] & 10/10 & .004 \\
Joint execution $-$ CQL, catching$^{b}$ & +0.034 [$-$0.010, 0.079] & 8/10 & .21 \\
8$\times$8, largest catching together & +0.016 [0.004, 0.028] & 9/10 & .60 \\
\bottomrule
\multicolumn{4}{@{}l}{$^{a}$IQL predators vs CQL prey, unlimited stamina. $^{b}$Faster prey, unlimited stamina.}
\end{tabular}
\end{table}

Joint execution was the one clear gain on 5$\times$5, and it did not carry over to 8$\times$8. When the prey were faster and stamina was unlimited, joint-execution CQL predators facing CQL prey beat their parts and flanked more (Table~\ref{tab:effects}), and caught the prey in 23.7 steps instead of 40.8 steps, a 42\% cut that held in all 10 seeds and with best moves only. Even then, a post-hoc check found that they caught prey together no more often than random predators, so they fell short of our full definition of coordination. Elsewhere on 5$\times$5, joint execution helped little or not at all, and with faster predators and limited stamina it slowed capture by 5.7 steps. On 8$\times$8 it helped nowhere. Its predators were 6 to 15\% slower than CQL predators in every setting and seed. In the four settings we replayed, they gained nothing from hunting together. With best moves only, they were somewhat faster with faster prey but still did not coordinate. Part of their slowdown in training therefore came from exploring as a team. The two grids also differ in how much of the joint table the team had tried. In the states the predators occupied, the joint-execution team had never tried its best joint action in 14 to 27\% of visits on 5$\times$5 and 80 to 94\% on 8$\times$8.

On 5$\times$5, speed and stamina changed coordination in the opposite direction to our predictions. Catching together, against random predators, was highest at equal speed, lower with faster predators and lowest with faster prey, in all eight combinations of pairing and stamina, and limited stamina lowered it in 11 of the 12 combinations of pairing and speed. Table~\ref{tab:verdicts} gives the remaining verdicts.

\subsection{Controls}

Without reward shaping, the self-play gap between CQL and IQL at equal speed shrank to at most 1.4 steps on both grids. The self-play difference between these methods thus depends on shaping. Training five times longer on 5$\times$5 (1,000,000 episodes, 3 seeds) roughly doubled the gap, so CQL did not catch up with more training. In a best-response test, new prey trained for 100,000 episodes against the frozen predators of IQL self-play. On 5$\times$5 with unlimited stamina they outlasted the original prey by 10.7 to 47.3 steps, so those predators could be exploited. Elsewhere the new prey were still improving, which leaves the test inconclusive. Finally, learning had levelled off: over the last 10,000 training episodes no pairing changed by more than 1.3 steps, except in one setting that Section 8 discusses.

\section{When a Tabular Comparison Misleads}
\label{sec:misleads}

In the main comparison of Section 5, both methods broke ties at random (Section 3). In the usual implementation, the two methods differ in one detail. IQL takes its greedy action with \texttt{argmax}, which picks the first action in the list when several tie. CQL breaks ties at random. On the 8$\times$8 grid, this detail decides which method wins, even though the agents have learned.

\begin{table}[t]
\caption{Self-play capture time, CQL minus IQL, in steps (positive: CQL slower). IQL breaks ties by the first action (usual) or at random, as CQL does (matched). $^{*}$: significant after Holm correction over the six settings (10 seeds, paired by seed).}
\label{tab:ties}
\small
\begin{tabular}{@{}lrrrr@{}}
\toprule
 & \multicolumn{2}{c}{8$\times$8} & \multicolumn{2}{c}{5$\times$5} \\
Setting & usual & matched & usual & matched \\
\midrule
Equal speed & +5.5$^{*}$ & $-$5.6$^{*}$ & +7.2$^{*}$ & +7.6$^{*}$ \\
Equal speed, limited stamina & +6.4$^{*}$ & $-$10.9$^{*}$ & +1.6$^{*}$ & +3.0$^{*}$ \\
Faster predators & +8.4$^{*}$ & +0.4 & $-$5.5 & $-$4.2 \\
Faster predators, limited stamina & +6.2$^{*}$ & $-$5.8$^{*}$ & $-$9.1$^{*}$ & $-$8.3$^{*}$ \\
Faster prey & +0.2 & $-$6.2$^{*}$ & $-$1.6 & $-$0.5 \\
Faster prey, limited stamina & +13.4$^{*}$ & $-$4.6$^{*}$ & $-$0.3 & +1.0$^{*}$ \\
\bottomrule
\end{tabular}
\end{table}

With IQL's default rule, IQL self-play was faster than CQL self-play in five of the six settings (Table~\ref{tab:ties}). When both methods broke ties at random, the result turned around: CQL self-play was faster in five settings. The winner swaps in four settings. In the other two, a difference appears or disappears. On 5$\times$5 the two versions differ by at most 2 steps (Section 5.2).

Why does a tie rule still decide the comparison after 200,000 episodes, when the agents passed the learning check? The answer is coverage (Table~\ref{tab:coverage}). In the states they occupied, the 8$\times$8 agents had never tried the action their table rated best in 51 to 82\% of visits for IQL and 87 to 94\% for CQL, against at most 27\% on 5$\times$5. As Section 2 noted, untried actions look best when values start at zero and most rewards are negative. These untried actions tie with each other, and the tie rule then picks the move. So most of the 8$\times$8 agents' moves were still chosen among untried actions. The effect is large. In every setting, episodes between two teams that break ties by the first action ended 6.5 to 18.0 steps sooner than episodes between two teams that break ties at random. Most of this came from the prey. The prey's tie rule alone moved capture time by 6 to 16 steps in five of the six settings. The first action in the list is a move to the right. Agents that share this rule tend to move the same way, and the rule becomes a convention between them \cite{boutilier1996planning,wei2016lenient}.

The same habit fakes coordination. We replayed the 8$\times$8 predators with their own tie rule. Predators that break ties by the first action showed more move dependence than predators that break ties at random, in all 12 combinations of method and setting. Yet they never coordinated: neither catching together nor the pair's gain passed our rule. On its own, move dependence would have reported teamwork where there was only a shared habit. Our definition of coordination (Section 4) guards against this. It also requires catching together or flanking, and the pair beating its parts.

Three checks show when we can trust a tabular comparison of learning methods. The first plays trained agents and random agents against the same random opponents, as in Section 5.1. The second measures coverage. It counts how much of the state and joint-action space the agents tried and how often they had never tried the action they rated best (Table~\ref{tab:coverage}). The third gives the compared methods the same tie rule and reports it. The first check follows established advice \cite{tan1993multi,bellemare2013arcade,patterson2024empirical}. The second check, coverage, is the one our case shows we need. On 8$\times$8 the learning check passed. Yet the coverage check showed that most moves were still chosen among untried actions. And a coding default decided which method won. The third check follows from a finding noted in Section 2: implementation details can decide a comparison.

\section{Discussion}
\label{sec:discussion}

CQL cost the agents in both roles, and its one benefit came from how it acted (Section 5). Acting on the best joint action helped on 5$\times$5 with faster prey but slowed the predators on 8$\times$8. A plausible reason, which we did not test directly, is how much of the joint table the team had tried. Untried joint actions look best, for the reason given in Section 2. On 5$\times$5 the table was well filled, so the best joint action was usually one the team had learned. On 8$\times$8 the team had almost never tried it, so acting on it amounted to trying new moves. The table's size points the same way: CQL must learn 25 joint actions in each state where an IQL agent learns 5, and it tried a much smaller share of them (Section 5.1).

Averaging over the partner's actions has known costs of its own. Poor joint actions pull down the values of good ones, an effect called action shadowing \cite{fulda2007predicting}. Averaging over a partner who might do anything also favours the safe choice over the coordinated one \cite{christianos2023pareto}. Our results suggest a condition on the alternative: acting on the joint action may help only where the team has tried the joint actions. This agrees with a recent comparison in which a centralized controller found better solutions than independent learners when it had ample experience but lost that edge in larger settings \cite{dekock2025exponentially}. It also agrees with the argument that cooperative value-based methods undervalue joint actions that exploration samples too rarely \cite{zhang2025optimistic}.

Self-play rankings call for care. Because self-play followed the larger role cost (Section 5.2), speed reversed which method looked better on 5$\times$5 but not on 8$\times$8. A study run only in self-play would conclude that the agents' bodies decide which method is better on one grid and not on the other. Yet the methods' costs in each role pointed the same way throughout. Comparing the methods role by role, as cross-play allows, avoids that error.

The agents' bodies mattered, though less simply than we expected. On 5$\times$5 they shifted the balance between the two role costs, which produced the reversal. They also changed coordination in the opposite direction to our predictions (Section 5.3). Following work on collaborative hunting \cite{tsutsui2024collaborative}, we had expected predators to coordinate more when slower than the prey. Instead they caught prey together least then, and limited stamina lowered it further. Only the joint-execution gain came where we expected help, with faster prey. On 8$\times$8 the bodies changed the size of CQL's costs but not their direction.

Trained predators ganged up without coordinating. On 5$\times$5 they ended up together at a capture far more often than random predators (Section 5.3). Yet they approached from the same side, and the pair seldom beat its parts. Two predators that each chase the nearest prey would produce this pattern. Each predator faces a choice shaped like a social dilemma: rush the prey alone, or hold back and flank, which can help the team but costs the one who waits. The choice resembles the lone-wolf and pack captures of the Wolfpack game \cite{leibo2017sequential} and the Stag Hunt, in which hunting together pays more and hunting alone is safer \cite{christianos2023pareto}. We did not measure these payoffs. On 8$\times$8 there was no coordination of any kind to explain.

The question of how to act from joint values also arises in centralized training with decentralized execution, where each agent acts on its own. As Section 2 noted, value-factorization methods ask when each agent's greedy action matches the greedy joint action \cite{son2019qtran}. Linear factorization in effect averages over partners, weighted by how often their actions appear in the data \cite{wang2021towards}. Our results suggest testing the way of acting explicitly there too, together with how much of the joint-action space the agents have tried. In the language of equilibria, averaging settled for a worse outcome in which each predator chased alone. Acting on the joint action escaped it only in one setting of the small grid. Failing to select the better of several equilibria is a known cause of miscoordination among independent learners \cite{matignon2012independent}. Exact ties are rare with neural networks, so the tie artefact of Section 6 may not carry over, but thin coverage of a large joint-action space may.

Reward design also changes such comparisons. Without shaping, the self-play gap between the methods shrank or vanished on both grids (Section 5.4). This fits the result that potential-based shaping leaves the set of best trade-offs in a multi-agent problem unchanged but can change which equilibrium the agents reach \cite{mannion2017policy}.

\section{Limitations}
\label{sec:limitations}

Our study has limits of task, measurement and training. The task never requires teamwork, because one predator can capture a prey alone. We implemented a variant that needs two predators per capture but did not run it. Our learners are tabular, with two agents per team on two small grids, so the results may not hold for larger teams, larger state spaces or learners based on neural networks. Two grid sizes do not make a trend, but they show which results held as the grid grew. CQL's costs in both roles, the lack of coordination and the self-play rule of Section 5.2 held on both grids, and we expect them to hold on larger grids. The joint-execution gain, found in one setting of the small grid, vanished on 8$\times$8, and the tie rule came to decide the winner. Both appear to depend on coverage (Sections 6 and 7). At the same budget a larger grid leaves more untried, so we expect such effects to grow. If the budget grew with the grid, larger grids might or might not behave like 5$\times$5. We have tested none of these expectations.

Two measures are weaker than we planned. Move dependence rises whenever both predators react to the same situation (Section 4). The capture measures compare trained predators with random ones, which often end up on opposite sides of a prey by chance, so they may miss teamwork of another form. With 10 seeds, only effects shared by nearly all seeds can pass after correction, so our test is conservative.

In training, every method starts every value at zero, and most rewards are negative. The coverage and tie effects of Sections 5 and 6 depend on this, and we did not test other starting values. As Section 3 notes, the limited-stamina settings are partially observable, the update bootstraps after the final capture, and we did not tune the learning parameters for each method. Each applies to all methods, but its effect may differ between them, and we did not measure it. Several ablations covered one direction or setting only. IQL with team reward ran only in self-play, so its effect on each role is unknown. It uses CQL's summed reward, so the joint-table step keeps the reward scale and the shared-reward step changes it. We trained without shaping at equal speed only, and we replayed joint execution, tried only for CQL predators, in four of the six settings. The best-response test, for new prey only, was inconclusive with limited stamina and on 8$\times$8, where 100,000 episodes were too few. On 8$\times$8 with faster prey and unlimited stamina, learning had not levelled off. Episodes in most pairings still lengthened by 7 to 12 steps over the last 50,000 episodes, and the 200-step limit, reached by up to 21\% of episodes, understates the largest differences there. We did not save the tables at 1,000,000 episodes, so we do not know which team drives the wider self-play gap there. And on 5$\times$5 with unlimited stamina, we read the shaping comparison before learning without shaping had levelled off.

\section{Conclusion and Future Work}
\label{sec:conclusion}

We asked whether centralized learning helps embodied agents coordinate. To make the answer trustworthy, we also asked when a tabular study can tell. On two grids at the same budget, in a task where one predator can capture prey alone, central Q-learning was worse than independent Q-learning in both roles, and its predators coordinated no more than independent ones (Section 5). Its one benefit, acting on the best joint action, sped up captures only on the small grid, with faster prey and unlimited stamina. In self-play, the larger of the two role costs decided which method looked better. On the larger grid, a tie-breaking default still decided which method won after the agents had learned. Three checks show when we can trust a tabular comparison (Section 6): playing trained and random agents against the same random opponents, measuring state-space coverage and matching tie-breaking.

Our results matter most to researchers who compare multi-agent learning methods, to those who design learning for robots and other embodied teams, and to those who use learning agents to study coordination and social behaviour. We cannot judge whether centralized learning helps apart from how a centralized learner acts and how much of the joint-action space it has tried. Claims about coordination need direct measures: here predators often caught prey together without the pair beating its parts, and a shared tie habit made their moves look coupled. For embodied teams, limits on speed and energy changed the size and balance of each method's costs, so learning methods should be compared under the team's own limits. Any comparison, including studies of coordination and social behaviour \cite{leibo2017sequential,leibo2021meltingpot}, can find a large, significant and repeatable difference that comes from an implementation default when experience is thin. The checks we propose are cheap. Finally, the study suits students and early researchers entering multi-agent learning. A tabular setting has no function approximation, so every value can be inspected and a result traced to its cause, as we traced the tie effect in Section 6. Our checks show which details still need control even in this simple setting.

Several directions follow. A task in which one predator cannot capture alone would test whether centralization helps when teamwork is required. A multi-objective view would treat each predator's captures, the team's capture speed and the energy spent as separate objectives and ask which trade-offs each method reaches \cite{radulescu2020momadm,hayes2022practical}. In deep multi-agent learning, averaging over partners and acting on the joint action deserve a direct comparison within value-factorization methods, as Section 7 suggests. Finally, other starting values and longer runs with the tables saved would settle some open cases of Section 8.

\begin{acks}
This work was supported by an internal research grant from Habib University.
\end{acks}

\section*{Use of AI Tools}
We used an LLM assistant to polish the text and to help write the experiment code. The authors checked all content and take responsibility for it.

\balance
\bibliographystyle{ACM-Reference-Format}
\bibliography{refs}

\end{document}